# Ultra-Uniform Nanocrystalline Materials via Two-Step Sintering

Yanhao Dong[1], Hongbing Yang[2], Lin Zhang[3], Xingyu Li[3], Dong Ding[4], Xiaohui Wang[5], Ju Li[1,6,*], Jiangong Li[2,*], I-Wei Chen[7,*]

[1]*Department of Nuclear Science and Engineering, Massachusetts Institute of Technology, Cambridge, MA 02139, USA*

[2]*Institute of Materials Science and Engineering and MOE Key Laboratory for Special Functional Materials and Structure Design, Lanzhou University, Lanzhou 730000, China*

[3]*Beijing Advanced Innovation Center for Materials Genome Engineering, Institute for Advanced Materials and Technology, University of Science and Technology Beijing, Beijing 100083, China*

[4]*Energy & Environmental Science and Technology, Idaho National Laboratory, Idaho Falls, ID 83415, USA*

[5]*State Key Laboratory of New Ceramics and Fine Processing, School of Materials Science and Engineering, Tsinghua University, Beijing 100084, China*

[6]*Department of Materials Science and Engineering, Massachusetts Institute of Technology, Cambridge, MA 02139, USA*

[7]*Department of Materials Science and Engineering, University of Pennsylvania, Philadelphia, PA 19104, USA*

**Nanocrystalline metals and ceramics with <100 nm grain sizes and superior properties (e.g., mechanical strength, hardness, fracture toughness and stored dielectric energy) are of great interest. Much has been discussed about achieving nano grains, but little is known about maintaining grain-size uniformity that is critical for material reliability. An especially intriguing question is whether it is possible to achieve a size distribution narrower than what Hillert[1] theoretically predicted for normal grain growth, a possibility suggested—for growth with a higher growth exponent—by the generalized mean-field theory[2] of Lifshitz, Slyozov, Wagner (LSW)[3,4] and Hillert but never realized in practice. We demonstrate that this can be achieved in bulk materials with an appropriately designed two-step sintering route that (a) takes advantage of the large growth exponent in the intermediate sintering stage to form a most uniform microstructure despite porosity remaining, and (b) freezes the grain growth thereon while continuing densification to reach full density. The resultant dense bulk $Al_2O_3$ ceramic has an average grain size of 34 nm and a much narrower size distribution than Hillert's prediction. Bulk $Al_2O_3$ with a grain-size distribution narrower than the particle-size distribution of starting powders was also demonstrated using this strategy, as were highly uniform bulk engineering metals and ceramics of either high purity and high melting points (Mo and W-Re) or highly complex compositions (core-shell $BaTiO_3$ and $0.87BaTiO_3$-$0.13Bi(Zn_{2/3}(Nb_{0.85}Ta_{0.15})_{1/3})O_3$).**

Rapid development in the past decades has provided nanocrystalline materials with improved and emerging properties.[5-7] While the benefits of nano-structuring (e.g., grain-boundary strengthening via the Hall-Petch relationship) are mainly attributed to the average grain size, it is the size distribution that matters most to engineering reliability.[8,9] This is because mechanical, electrical, dielectric, and other failures dictated by instantaneous or gradual breakdown events typically happen at the weakest point, which is often associated with a microstructural or chemical inhomogeneity at some grain or grain boundary. Reducing grain size dispersion will make all grains alike, which tends to eliminate inhomogeneity. Therefore, it should be a major goal in nanomaterial development.

When all the grain boundaries have the same energy and mobility and cooperatively move to lower the total grain-boundary energy in a polycrystal, it leads to a parabolic "normal" growth law, $(\text{grain size})^2 \sim \text{time}$. According to Hillert[1], the theoretical ratio $\sigma$, which is the ratio of the standard deviation $\Sigma$ of the grain-size distribution to the average grain size $G_{avg}$, should be 0.354 in normal grain growth. Since real materials do not have grain boundaries of the same energy and mobility, they should have a larger $\sigma$ during grain growth, which is true for a large number of dense ceramics sintered in various ways[8,10-45] as shown by the blue data points in **Fig. 1**. Meanwhile, **Figure 1** also reveals that smaller $\sigma$ tends to come with a smaller $G_{avg}$. This trend explains why nanomaterials are attractive: they tend to combine small $G_{avg}$ and small $\sigma$, which is the key to exceptional properties and reliability. Thus, the state of the art suggests that there is a lower limit of grain-size dispersion set by Hillert's $\sigma$=0.354, and it would be difficult, if not impossible, to do better in real dense materials. But if a better nanomaterial is to come, one must challenge this limit and find a way to obtain bulk, dense engineering nanomaterials with a smaller $\sigma$. We will demonstrate that this is possible by taking advantage of a growth stagnation in intermediate-stage sintering.

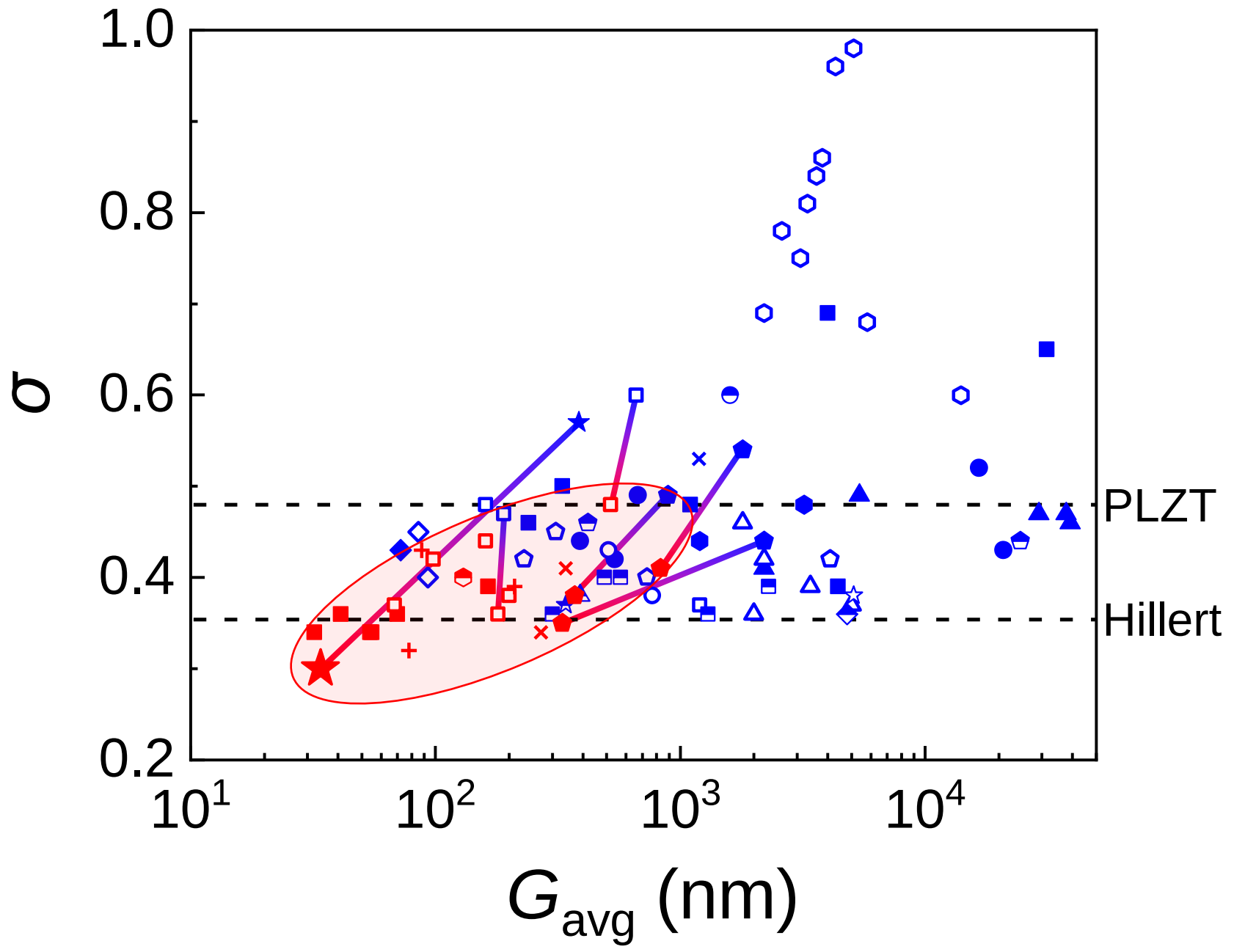

**Figure 1 "Quality plot" of $G_{avg}$ and $\sigma$ of sintered materials.** $G_{avg}$ and $\sigma$ of 25 dense bulk materials represented by 24 different types of symbols listed in **Supplementary Table S1**, variously sintered as reported in the literature and in this work. Two-step sintered in red, conventionally sintered in blue, and those by both methods (all in this work except for W) with the same powders and green body forming treatment are connected by solid lines. Lower dashed line: Hillert's theoretical prediction.[1] Upper dashed line: hot-pressed transparent (Pb,La)(Zr,Ti)$O_3$ (PLZT) with a known uniform microstructure.[33] Shaded ellipse is for guidance of eyes.

To motivate our approach, we recall that essentially mono-sized nanoparticles are routinely obtained by liquid-phase precipitation after proper growth and coarsening—also called Ostwald ripening, and this paradigm is relevant to bulk sintering because Ostwald ripening is described by the same LSW theory.[3,4,46,47] The common rationale is that particle-size evolution is a self-sharpening process due to the rapid decay of growth rate with size. This growth/coarsening stagnation is caused by (a) a diminishing driving force (capillary pressure) that inversely scales with the particle size and (b) a diminishing diffusion kinetics that has a diffusion distance proportional to the particle size. Mathematically, a faster approach to stagnation is signaled by a higher exponent of the growth law, (particle/grain size)$^n$ ~ time, where $n$ is 3 in Oswald ripening and 2 in normal grain growth. Inasmuch as $\sigma$ decreases with $n$, one stands to reason that if a higher growth exponent can somehow be engineered into sintering, then it would be possible to obtain a more uniform sintered microstructure than that predicted by Hillert. This will yield a better nanomaterial that lies closer to the lower-left corner of **Fig. 1**: one with "ultra-fine" and "ultra-uniform" grains. How to achieve it in theory and in practice is illustrated below for a dense bulk $Al_2O_3$ ceramic with $G_{avg}$=34 nm and $\sigma$=0.30, indicated by the red star in **Fig. 1**. Other small-$\sigma$ materials of practical interest obtained by the same method will also be described.

Our reasoning is supported by the analytic solution of steady-state size distribution for the LSW-Hillert growth equation[2],

$$\frac{dG}{dt} = 2M\gamma\left(\frac{G}{a}\right)^{\alpha}\left(\frac{1}{G_{cr}} - \frac{1}{G}\right) \qquad (1)$$

Here, $G$ is the size of a growing or shrinking grain/particle, $M$ is the mobility, $\gamma$ is the interfacial energy, $a$ is a length (e.g., atomic spacing) that preserves the dimension, $2\gamma/G$ is the capillary pressure, and $G_{cr}$ is the critical size of a grain/particle that neither grows nor shrinks at time $t$—thereby setting a chemical potential (in pressure) $2\gamma/G_{cr}$ for the system that ensures mass/volume conservation at $t$. More specifically, $M$ is the grain boundary mobility in normal grain growth, and $M=\frac{D}{k_B T}\frac{\Omega}{a}$ (with $D$ the diffusivity, $\Omega$ the atomic volume, $k_B$ the Boltzmann constant and $T$ the absolute temperature) in particle growth. The different growth dynamics signaled by $n$ comes from $\alpha$, which is 0 in normal grain growth and −1 in LSW (Ostwald ripening). For an arbitrary $\alpha \leqslant 1$, we obtained a steady-state power-law $G_{avg}^n \sim t$ with a growth exponent $n=2-\alpha\geq 1$, and a close form solution[2] of the normalized size distribution $P'(u)$ for $u=G/G_{cr}$

$$P'(u)=\frac{3u^{1-\alpha}}{u^{2-\alpha}-\frac{(2-\alpha)^{2-\alpha}}{(1-\alpha)^{1-\alpha}}(u-1)}\exp\left[-\frac{3}{2-\alpha}\int_0^u \frac{-(2-\alpha)u^{1-\alpha}}{\frac{(2-\alpha)^{2-\alpha}}{(1-\alpha)^{1-\alpha}}(u-1)-u^{2-\alpha}}du\right] \qquad (2)$$

A more convenient distribution $P(G/G_{avg})$ in terms of the normalized grain size $G/G_{avg}$ is obtained from $u=\frac{G}{G_{avg}}u_{avg}$, where $u_{avg}=\int_0^{u_0} uP'(u)du$ with $u_0$ being the upper cutoff of the distribution. **Figure 2** clearly illustrates that $P(G/G_{avg})$ sharpens as $\alpha$ decreases and $n$ increases; meanwhile, its $\sigma=\Sigma/G_{avg}$ (where $\Sigma$ is the standard deviation of $G$) decreases as shown in the inset of **Fig. 2.** Clearly, to beat Hillert's $\sigma$ =0.354 limit, one must find an $\alpha$<0 ($n$>2) process to replace normal grain growth.

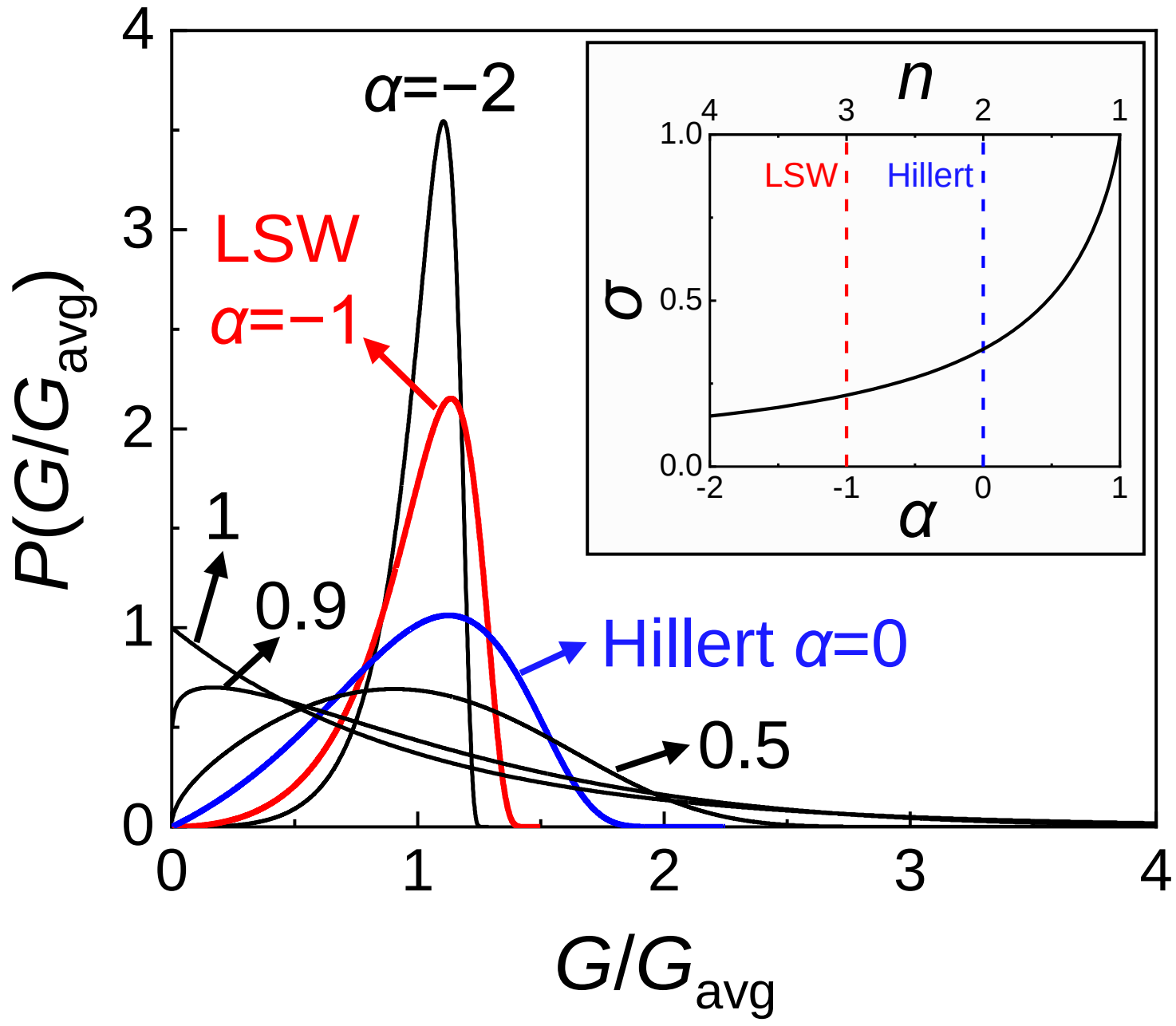

**Figure 2 Steady-state solution of generalized growth equation.** Calculated normalized grain size distribution $P(G/G_{avg})$ as a function of $\alpha$. Inset: Calculated standard deviations $\sigma$ for $G/G_{avg}$ as a function of $\alpha$ and $n$.

Normal grain growth with $n$=2 was firmly established in dense ceramics, e.g., in undoped or variously doped $Y_2O_3$, $CeO_2$ and $ZrO_2$.[20,21,38] However, it is known that the grain-growth exponent $n$ is ≥3 in porous ceramics. This is because the grain boundaries of a larger grain are statistically in contact with more pores, hence sense stronger pinning forces.[48] The latter applies to intermediate-stage sintering, defined as sintering of a powder compact in which already-neck-bonded particles are still separated by mostly open porosity (namely, interconnected pore channels) instead of isolated pores as in final-stage sintering. Therefore, we hypothesized that $\sigma$ of the intermediate-stage microstructure is smaller than Hillert's 0.354, and if the hypothesis is verified, we hoped such a state could provide a "template" to grow materials with unprecedented uniformity.

To verify the hypothesis, we sintered powder compacts made of 4.7 nm high-purity $Al_2O_3$ powders without applied pressure, i.e., by free sintering. After reaching a set temperature, the partially dense compact was immediately cooled and analyzed. (See **Methods** for experimental details.) Although a higher set temperature always led to a higher relative density $\rho$, $G_{avg}$ (open circles in **Fig. 3a**) and $\Sigma$ (error bar), both measured from the transmission electron microscopy (TEM) images of fractured fragments, the variation of $\sigma$ (open circles in **Fig. 3c**) is non-monotonic and reaches a minimum at $\rho$=65%. This density with the smallest $\sigma$ will be called $\rho_u$, where the subscript "u" stands for "uniform". The same trend was confirmed by the data in **Fig. 3b (**$G_{avg}$ as filled circles and $\Sigma$ as error bars) and **Fig. 3c** ($\sigma$ as filled circles) measured from scanning electron microscopy (SEM) images of fractured and thermally etched surfaces of the same set of samples. Since $\sigma$ during much of the $\rho$=60-85% range falls significantly below Hillert's $\sigma$=0.354 (dashed horizontal line in **Fig. 3c**), our hypothesis is confirmed.

Uniform intermediate-stage microstructure is evident from the micrograph and grain-size histogram of an $Al_2O_3$ sample with $\sigma$=0.31 at $\rho$=84%, shown in **Fig. 4a-b**, taken from the SEM images of fractured and thermally etched surfaces. Comparably small $\sigma$ values were also obtained in other free-sintered porous ceramics, 8 mol% yttria-stabilized cubic zirconia (8YSZ) with $\sigma$=0.30 (orange square in **Fig. 3c**, measured from the SEM images of polished and thermally etched surfaces) at $\rho$=69%≈$\rho_u$, and 3 mol% yttria-stabilized tetragonal zirconia (3YSZ) with $\sigma$=0.33 (purple triangle in **Fig. 3c**, measured from the SEM images of polished and thermally etched surfaces) at $\rho$=70%≈$\rho_u$. Their micrographs and grain size histograms (**Fig. 4c-d** for 8YSZ and **Fig. 4e-f** for 3YSZ) all portray a highly uniform microstructure despite the prominent presence of open pore channels.

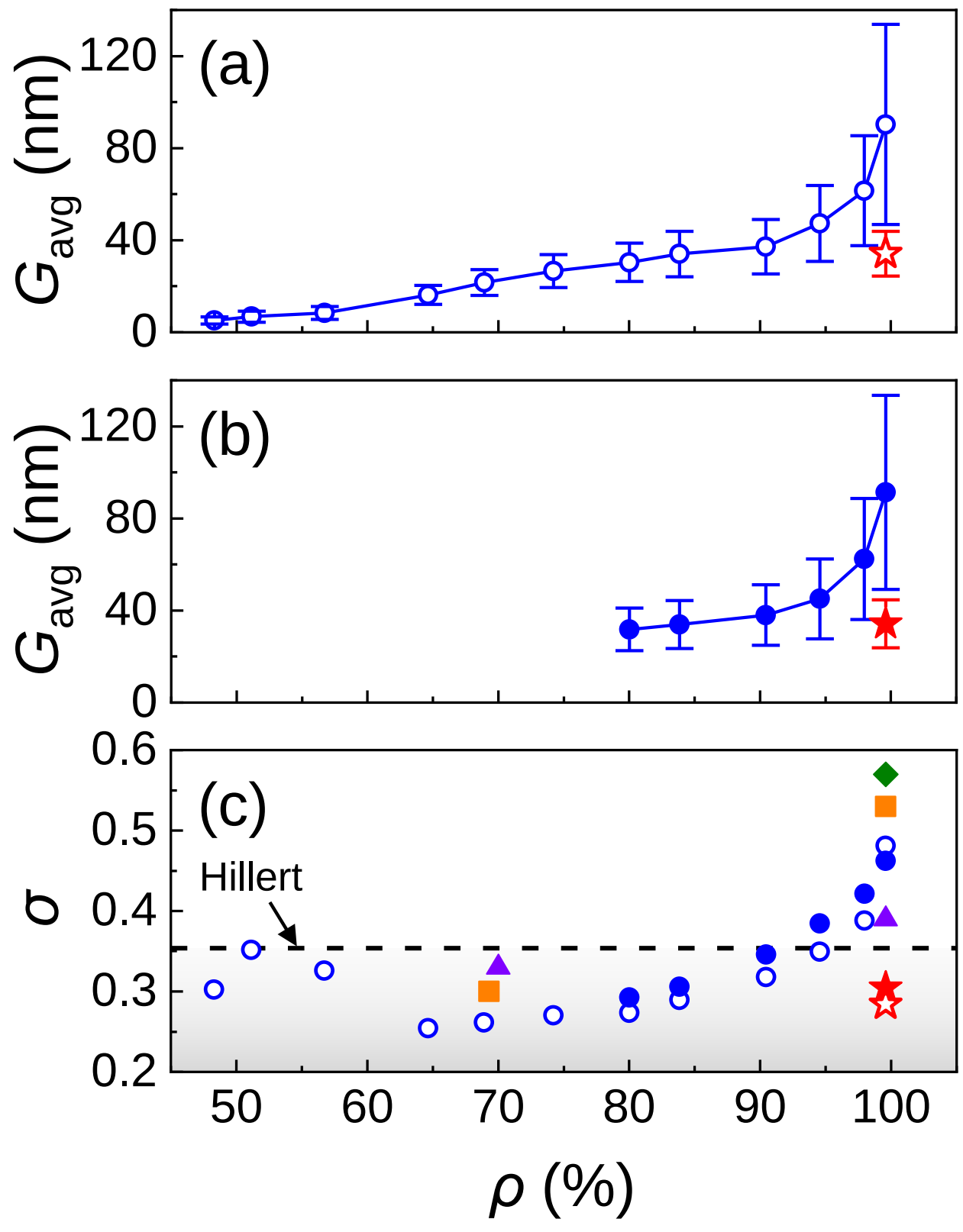

**Figure 3 Microstructural dispersity in porous and dense polycrystals.** Trajectory of average grain size $G_{avg}$ vs. relative density $\rho$ of $Al_2O_3$ ceramics measured under (a) TEM in open symbols and (b) SEM in closed symbols, with standard deviation $\Sigma$ shown as error bars. Two-step sintering data at full density also plotted as stars. (c) Relative standard deviation, $\sigma=\Sigma/G_{avg}$, from (a-b) using the same symbols. Also included are orange square (8YSZ) at $\rho$=69% and purple triangle (3YSZ) at $\rho$=70%;

orange square (8YSZ), purple triangle (3YSZ), and green diamond ($Al_2O_3$) at $\rho$≈100%: the latter samples annealed for additional time at highest temperature after reaching nearly full density. (See text for details.)

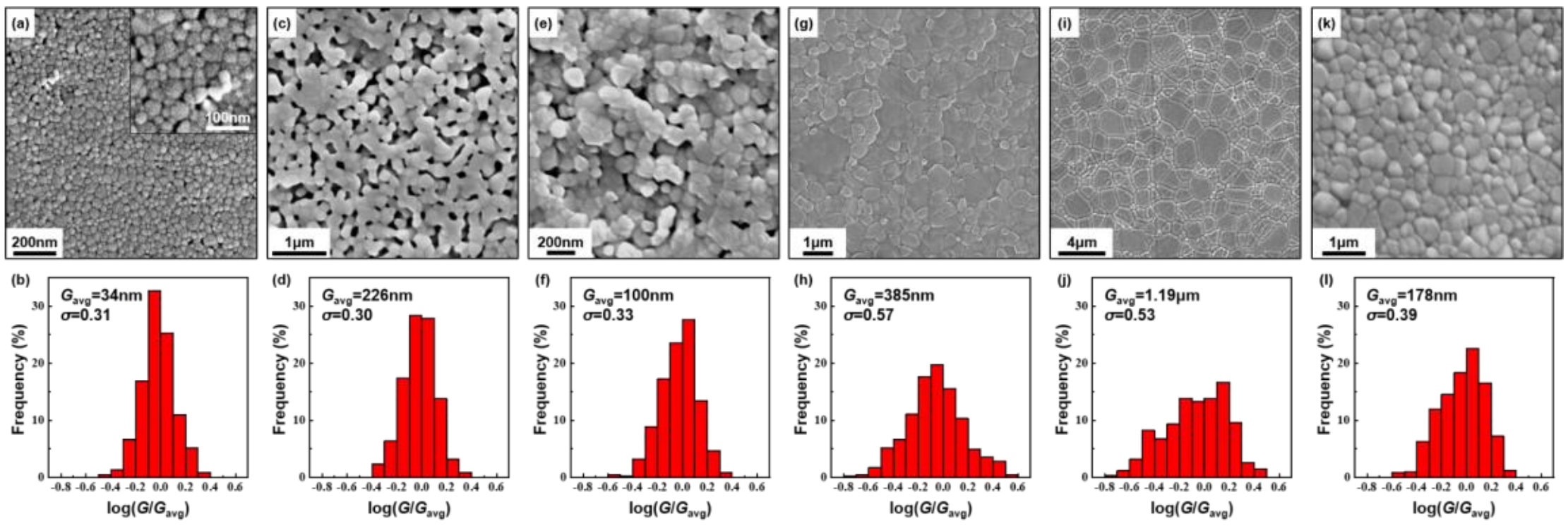

**Figure 4 Comparing microstructural uniformity in porous and dense ceramics.** SEM micrographs and normalized grain size histograms of (a-b) $Al_2O_3$ heated to 1150$^o$C without holding, $\rho$=84%; inset in (a): enlarged section showing porosity; (c-d) 8YSZ heated to 1280$^o$C without holding, $\rho$=69%; (e-f) 3YSZ heated to 1240$^o$C without holding, $\rho$=70%; (g-h) dense $Al_2O_3$ sintered at 1300$^o$C for 10 h; (i-j) dense 8YSZ sintered at 1300$^o$C for 12 h; and (k-l) dense 3YSZ sintered at 1300$^o$C for 12 h. Also listed in histograms are $G_{avg}$ and $\sigma$.

Unfortunately, although a remarkably small $\sigma$ throughout the intermediate-stage sintering is now established, the uniformity invariably deteriorates as most open porosity is eliminated—especially when density exceeds 90% (see **Fig. 3c**) that marks the onset of final-stage sintering in most materials.[49,50] The common practice of holding the material at the highest sintering temperature for additional time to assure full densification makes matters worse, as it causes concurrent grain growth in nearly dense materials that increases $\sigma$ further. For example, $\sigma$ reached 0.57 in an $Al_2O_3$ ceramic (measured from the SEM images of fractured and thermally etched surfaces) held for 10 h at 1300°C as indicated by the green diamond at $\rho$=100% in **Fig. 3c**. It has a visibly deteriorated microstructure (**Fig. 4g**) and a much less uniform grain-size histogram (**Fig. 4h**) than its $\rho$=84% predecessor (**Fig. 4a-b**). The same holds for 8YSZ and 3YSZ held at 1300°C for 12 h: they feature much larger $\sigma$ (0.53 for 8YSZ, orange square at $\rho$≈100% in **Fig. 3c**; 0.39 for 3YSZ, purple triangle at $\rho$≈100% in **Fig. 3c**; both measured from the SEM images of polished and thermally etched surfaces) and visibly worse microstructures (**Fig. 4i-l**) than their porous counterparts in **Fig. 4c-f**. Among them, 3YSZ experienced the least deterioration after high-temperature holding. This is because it is endowed with a strong solute drag that increases grain-growth resistance (which is why it is the first and best superplastic ceramic[51]). Yet its $\sigma$ still increased from 0.33 to 0.39 as $G_{avg}$ grew from 100 nm at $\rho$=70% to 178 nm after holding at 1300°C (**Fig. 4f&l**).

Clearly, a different sintering strategy is needed to retain the uniform intermediate-stage microstructure during densification. Two-step sintering[52] has been shown to do exactly that: it effects densification (via diffusion along stationary grain boundaries, which requires a relatively small activation energy) without grain growth (via grain-boundary migration that must perturb the grain-boundary network, which requires a relatively large activation energy). In this way, two-step sintering keeps the microstructure frozen while improving the density, and in the past 20 years it has been successfully demonstrated in numerous oxides, nitrides, carbides and some metals.[8,15-17,36,37,43-45,52-56] In practice, it first heats the powder compact to a higher temperature (designated as $T_1$) without holding to attain a high enough density, then lowers the temperature to $T_2$ and holds there for an extended time to reach full density. To succeed, the post-$T_1$ density should be high enough to render all remaining pores/pore channels unstable, which requires their surface curvatures to change from negative to positive (a spherical pore has a uniformly positive surface curvature of 2/radius). With 4.7 nm $Al_2O_3$ powders ($G_{avg}$=4.7 nm and $\sigma$=0.23 measured from the TEM images of dispersed powders; TEM micrograph and normalized grain size histogram in **Fig. 5a-b**), we used $T_1$=1150°C to reach $\rho$=84%, then cooled the sample down to $T_2$=1025°C with an additional hold for 40 h to obtain a final density of 99.6%. As shown by the microstructure in **Fig. 5c** and the grain size histogram in **Fig. 5d**, the dense body has essentially the same $G_{avg}$ (34 nm, measured from the SEM images of fractured and thermally etched surfaces; also shown in **Fig. 3a-b** by stars, open ones from TEM and filled ones from SEM) and $\sigma$ (0.30 from SEM; shown in **Fig. 3c** from red error bars in **Fig. 3a-b**) as those of the $\rho$=84% sample (post-$T_1$ and before cooling down to $T_2$, $G_{avg}$=34 nm and $\sigma$=0.30, **Fig. 4a-b**). Therefore, the second step not only arrests grain growth, but also preserves grain-size uniformity. Having successfully frozen the microstructure throughout $T_2$-sintering, we obtained a dense nanocrystalline bulk ceramic—the red star in **Fig. 1**—an “ultra-uniform” alumina, thus called because its $\sigma$=0.30 lies below Hillert’s limit.

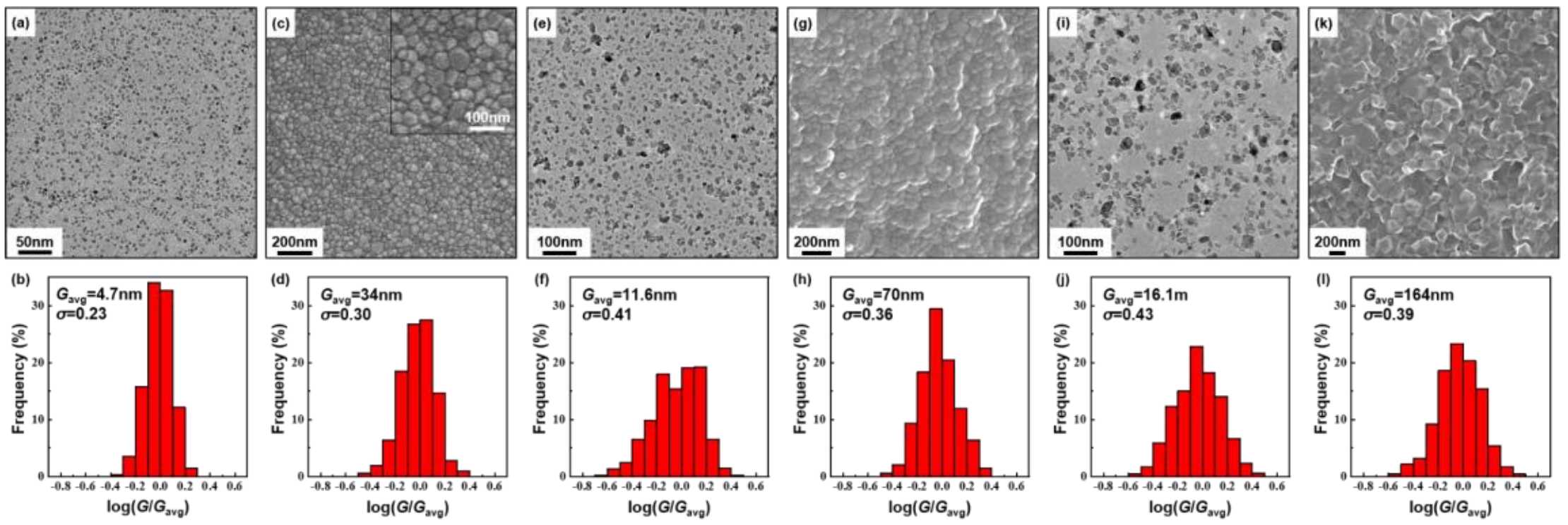

**Figure 5 Two-step sintered nanocrystalline bulk $Al_2O_3$ with different starting powders.** Micrographs and normalized grain size histograms of (a-b) 4.7 nm $Al_2O_3$ powders and (c-d) dense two-step sintered $Al_2O_3$ reaching 1150°C ($T_1$) without holding and then held at 1025°C ($T_2$) for 40 h using 4.7 nm $Al_2O_3$ powders; inset in (c): enlarged section showing absence of porosity. Corresponding micrographs and histograms of (e-f) 11.6 nm $Al_2O_3$ powders and (g-h) dense two-step sintered $Al_2O_3$ reaching 1200°C ($T_1$) without holding and then held at 1000°C ($T_2$) for 20 h using 11.6 nm $Al_2O_3$ powders. Similar micrographs and histograms of (i-j) 16.1 nm $Al_2O_3$ powders and (k-l) dense two-step sintered $Al_2O_3$ reaching 1275°C ($T_1$) without holding and then held at 1075°C ($T_2$) for 20 h using 16.1 nm $Al_2O_3$ powders. Also listed in histograms are $G_{avg}$ and $\sigma$.

Two-step free sintering also led to other bulk materials more uniform than possible by conventional sintering. As shown in **Fig. 6a-b,** two-step sintered bulk Mo ($T_1$=1180°C for 1 h, $T_2$=1110°C for 10 h) has a finer and more uniform microstructure ($G_{avg}$=370 nm and $\sigma$=0.38) than conventionally sintered Mo (1400°C for 3 h) in **Fig. 6c-d** that has $G_{avg}$=890 nm and $\sigma$=0.49. Likewise, two-step sintered 90W-10Re alloy ($T_1$=1200$^o$C for 1 h, $T_2$=1100$^o$C for 20 h) shown in **Fig. 6e-f** has a finer and more uniform microstructure ($G_{avg}$=330 nm and $\sigma$=0.35) than conventionally sintered 90W-10Re (1500$^o$C for 2 h) in **Fig. 6g-h** that has $G_{avg}$=2.2 μm and $\sigma$=0.44. These examples demonstrate that two-step free sintering can obtain bulk metals and alloys of extremely high melting points (2623$^o$C for Mo and ~3200$^o$C for 90W-10Re). This is especially note-worthy because it used unusually low sintering temperatures (≤1200$^o$C) and no low-melting-point sintering aid, which would have caused possible contamination and deleterious effects on properties.

Two-step free sintering was further used to densify complex perovskites derived from $BaTiO_3$, which is the basis of commercially important ceramic capacitors. As shown in **Fig. 6i-j**, the microstructure of two-step sintered $BaTiO_3$ ($T_1$=1250$^o$C for 1 min, $T_2$=1050$^o$C for 5 h) is more uniform ($G_{avg}$=180 nm and $\sigma$=0.36) than that of conventionally sintered one (1210$^o$C for 2 h, $G_{avg}$=190 nm and $\sigma$=0.47, **Fig. 6k-l**), the latter having some abnormally large grains (**Fig. 6k**) not found in **Fig. 6i**. Such uniformity is directly correlated to a 4× better retention time in field-holding tests at 185$^o$C for the two-step sintered ceramic.[8] Note that the $BaTiO_3$ powder used here was precoated with mixed dopant oxides of Ca, Ba, Y, Mg, Si, Mn and Ho to form a core-shell structure, which is needed to suppress leakage current and to smooth out $BaTiO_3$'s ferroelectric transitions at 120$^o$C and 8$^o$C, thus providing a broad, temperature-insensitive and low-loss dielectric response required for multilayer ceramic capacitor (MLCC) applications. By suppressing the migration of grain boundaries that would have availed their fast-diffusing paths to dopant interdiffusion thus compromised the core-shell structure, two-step sintering can better maintain property advantages. Another perovskite of $0.87BaTiO_3$-$0.13Bi(Zn_{2/3}(Nb_{0.85}Ta_{0.15})_{1/3})O_3$ (BT-BZNT) composition was also two-step sintered ($T_1$=1190°C for 1 min, $T_2$=1040°C for 3 h) to obtain a microstructure ($G_{avg}$=520 nm and $\sigma$=0.48, **Fig. 6m-n**) more uniform than obtained from conventional sintering (1150°C for 3 h, $G_{avg}$=660 nm and $\sigma$=0.60, **Fig. 6o-p**), where the latter also resulted in some abnormal grain growth not seen in the former. The better uniformity is directly correlated to a higher electrical field breakdown strength, which allows it to be fabricated into MLCCs (by the same two-step sintering procedure) that stored a record-high density of dielectric energy.[57]

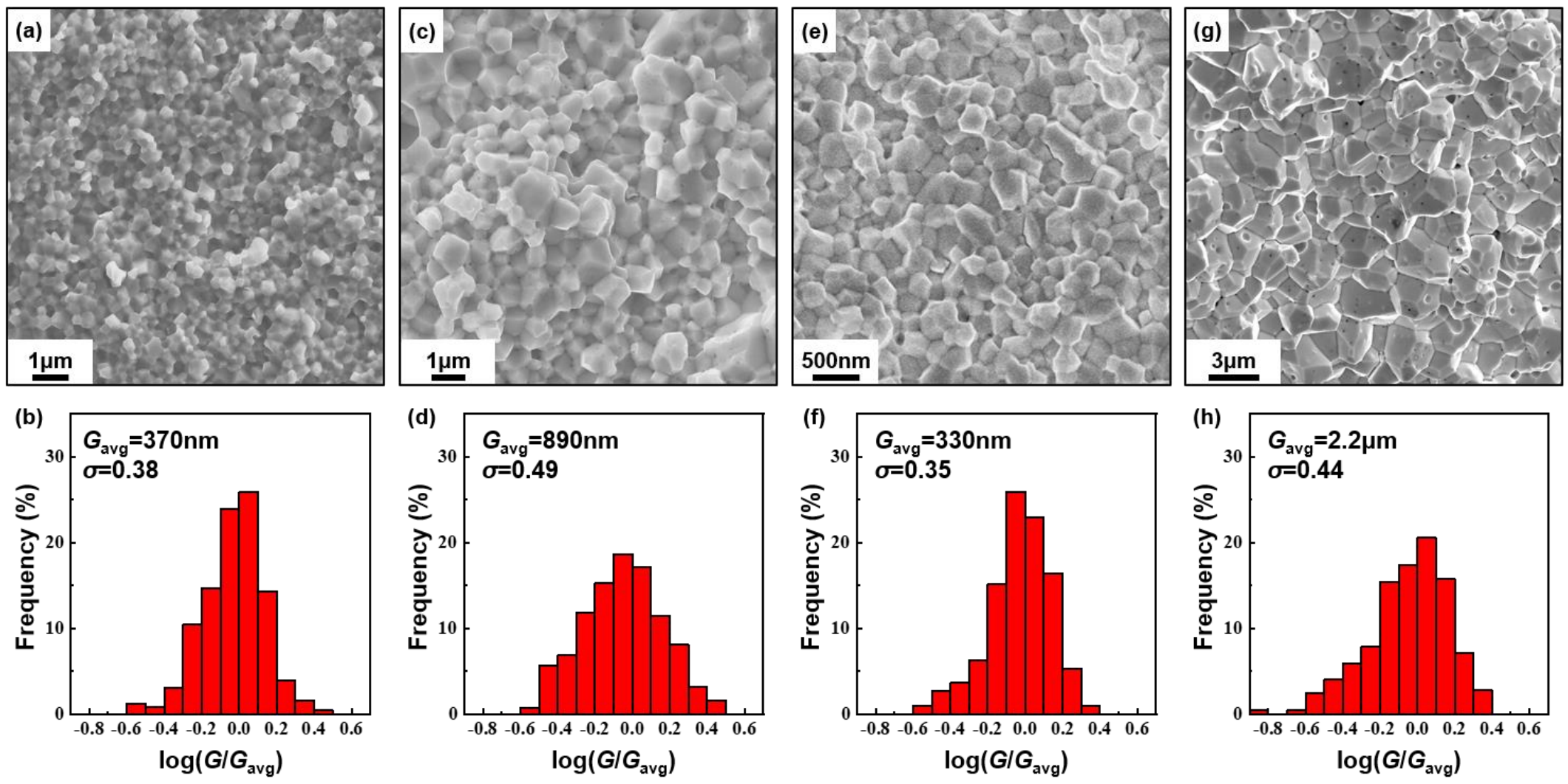

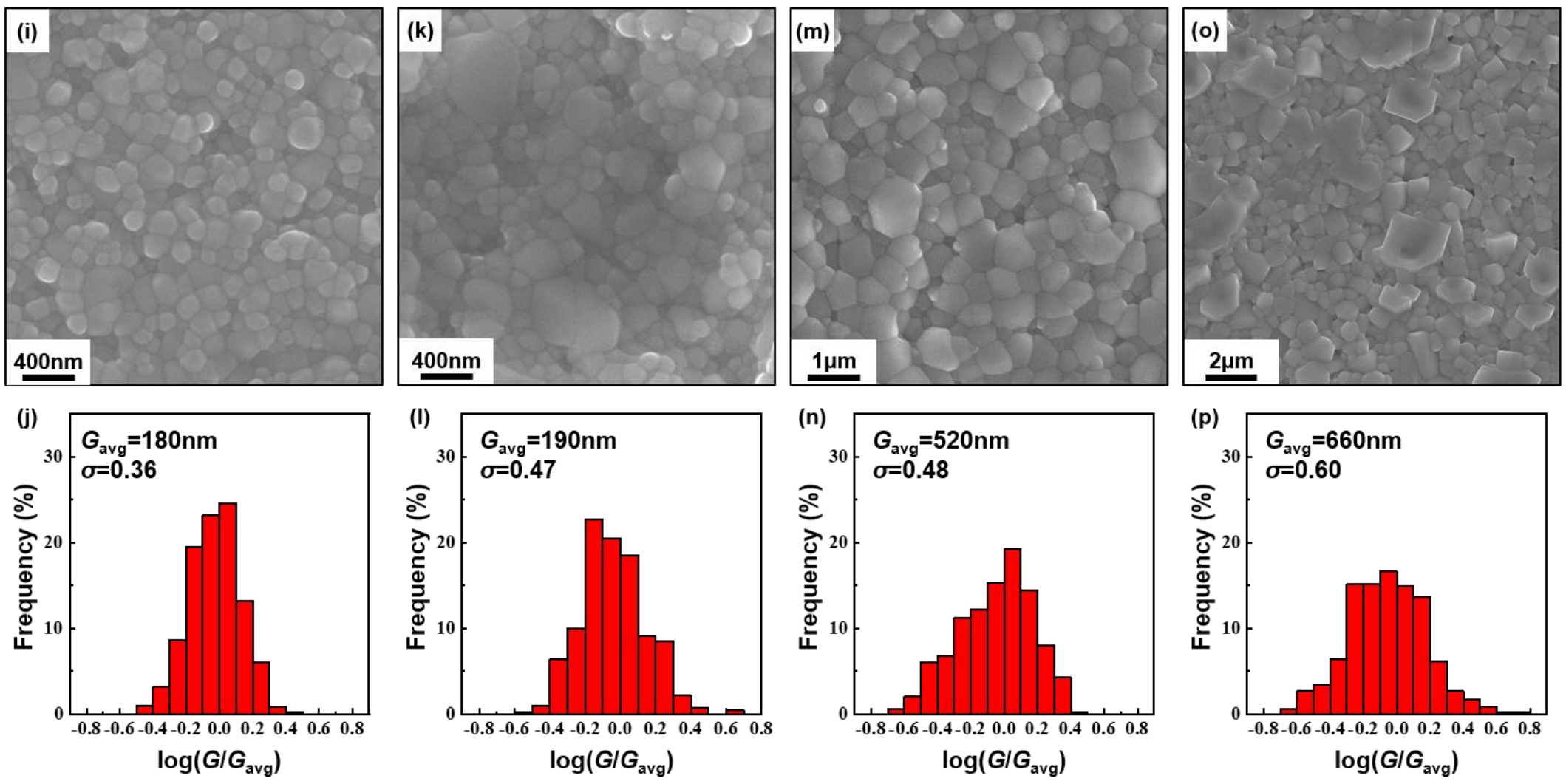

**Figure 6 Two-step free-sintered fine-grained bulk refractory metals and perovskite capacitors with uniform microstructure.** Micrographs and normalized grain size histograms of (a-b) dense two-step sintered Mo held at 1180°C ($T_1$) for 1 h and then held at 1110°C ($T_2$) for 10 h, compared to (c-d) Mo conventionally sintered at 1400°C for 3 h; (e-f) dense two-step sintered 90W-10Re held at 1200°C ($T_1$) for 1 h and then held at 1100°C ($T_2$) for 20 h, compared to (g-h) 90W-10Re conventionally sintered at 1500°C for 2 h; (i-j) dense two-step sintered $BaTiO_3$ (BT) held at 1250°C ($T_1$) for 1 min and then held at 1050°C ($T_2$) for 5 h, compared to (k-l) conventionally sintered BT sintered at 1210°C for 2 h; (m-n) dense two-step sintered $0.87BaTiO_3$-$0.13Bi(Zn_{2/3}(Nb_{0.85}Ta_{0.15})_{1/3})O_3$ (BT-BZNT) held at 1190°C ($T_1$) for 1 min and then held at 1040°C ($T_2$) for 3 h, compared to (o-p) conventionally sintered BT-BZNT at 1150°C for 3 h. Also listed in histograms are $G_{avg}$ and $\sigma$.

Previous sintering studies of density and pore/particle/grain-size evolutions found that $\sigma$ of open pores, as well as the size ratio of pore-to-grain, decrease rapidly during initial-stage sintering.[58,59] This was attributed to powder coarsening and neck formation, driven by surface diffusion, which in turn triggers particle relocation leading to homogenization of microstructure. These processes end when the powder compact becomes locked, which for spheres of the same size happens at $\rho_{rcp}$=0.63—the density of random close-packing—and usually at a somewhat higher density if particle sizes vary. Still driven by surface diffusion, further grain coarsening relative to pores eventually renders all pores thermodynamically unstable, which occurs at a higher density $\rho_c$. After $\rho_c$, grain-boundary diffusion alone can drive densification all the way to full density, albeit at the cost of increasing $\sigma$ of relative grain size unless two-step sintering is practiced. We suggest that particle/grain homogenization and locking can also explain why $\sigma$ of relative grain size reaches a minimum in **Fig. 3**, which occurs at $\rho_u$=0.65 that is just slightly above $\rho_{rcp}$. If so, homogenization during initial-stage sintering followed by microstructure freezing during two-step sintering will allow a dense microstructure with a relatively small final $\sigma$, even if the starting powder compact has a relatively large $\sigma$.

To test this idea, we used an $Al_2O_3$ powder ($G_{avg}$=11.6 nm and $\sigma$=0.41, measured from the TEM images of dispersed powders; TEM micrograph and normalized grain size histogram in **Fig. 5e-f**) less uniform than the one in the above example of $Al_2O_3$ ($\sigma$=0.23 for the powder) to sinter an intermediate sample of $\rho$=84% at $T_1$=1200$^o$C and a final sample of $\rho$=99.1% with $G_{avg}$=70 nm and $\sigma$=0.36 (all measured from the SEM images of fractured and thermally etched surfaces; microstructure and normalized grain size histogram in **Fig. 5g-h**) after holding at $T_2$=1000$^o$C for 20 h. Similarly, with another $Al_2O_3$ powder ($G_{avg}$=16.1 nm, and $\sigma$=0.43 measured from the TEM images of dispersed powders; TEM micrograph and normalized grain size histogram in **Fig. 5i-j**), we used $T_1$=1275$^o$C to reach $\rho$=87%, and $T_2$=1075$^o$C for 20 h to obtain a 98.9% dense $Al_2O_3$ with $G_{avg}$=164 nm and $\sigma$=0.39 (measured from the SEM images of fractured and thermally etched surfaces; microstructure and normalized grain size histogram in **Fig. 5k-l**). These ceramics having a final $\sigma$ smaller than powder's initial $\sigma$ support the idea that the $\sigma$ evolutions of the relative grain size and relative pore size are a twin-measure of the homogenization process throughout initial- and intermediate-stage sintering[58,59], and they reach their lowest at $\rho_u \approx \rho_{rcp}$ when the porous compact becomes locked. After that two-step sintering must be used to preserve the uniform microstructure to higher density.

These findings can be used to guide best practices for two-step sintering. Existing data suggest that $\rho_c$, which signals a critical ratio of grain size to pore size above which pore shrinkage is thermodynamically feasible without any grain-boundary movement, is similar for all cubic materials: ~75% in $Y_2O_3$[52, 53], ~73% in $BaTiO_3$[17], ~76% in $Ni_{0.2}Cu_{0.2}Zn_{0.6}Fe_2O_4$[17]. In contrast, $Al_2O_3$ is hexagonal with highly anisotropic interfacial energies, which is reflected in a dihedral-angle distribution from 76$^o$ to 166$^o$ (although this can be partially ameliorated by doping)[60] that is much more dispersed than seen in $Y_2O_3$ (88$^o$-132$^o$)[58]. So a much higher $\rho_c$ of ~83%[16, 56] is needed in $Al_2O_3$ to render pores with the least favorable dihedral angles thermodynamically unstable. Since $\rho_c$ already exceeds $\rho_u$ (where $G_{avg}$ is smaller and $\sigma$ is the smallest), going from $\rho_u$ to $\rho_c$ before $T_2$-sintering means more grain coarsening, and this must go the furthest in $Al_2O_3$ because it has the highest $\rho_c$. Indeed, we found the ratio of final $G_{avg}$ to initial powder size ranges from 6 to 10 for $Al_2O_3$, compared to 3-6 for Mo and 90W-10Re and less than 2 in $BaTiO_3$-based ceramics, indicating $Al_2O_3$ is intrinsically more difficult to sinter. Clearly, the selection of $T_1$ must pay close attention to this aspect in order to best capture $\rho_c$ and minimize grain coarsening before $T_2$-sintering. Meanwhile, attention also needs to be paid to powder agglomeration and non-uniform packing because they can give rise to lower local density even after an overall $\rho_c$ is reached, thus preventing densification during $T_2$-sintering.

Turning to $T_2$-sintering, $T_2$ has been empirically found by trial and error in the literature (typically 100-200°C below $T_1$) to allow the densification of many materials without grain growth. However, in 8YSZ we have recently identified, for the first time, a transition temperature $T_{tr}$: 2-grain-controlled boundary migration above $T_{tr}$, and 3/4-grain-junction-controlled boundary migration below $T_{tr}$.[45] Moreover, there is an unphysically high activation energy (10.8 eV) for 3/4-grain-junction-controlled boundary migration that suggests more and more grain junctions become frozen at lower and lower temperature, so the few remaining mobile boundaries will also become immobile if there are pores to pin them. Therefore, it is advantageous to choose a $T_2$ as low as feasible, which is 100°C below $T_{tr}$ (1300°C) in our 8YSZ study. On the other hand, once pores are all gone, the few remaining mobile boundaries can start migrating. Therefore, further prolong annealing will cause localized grain growth/shrinkage, which broadens the grain-size distribution. Again, a lower $T_2$ is advisable as it leaves a wider time window to terminate heating before microstructure begins deteriorating. (In 8YSZ, it takes 500-1000 h at ≤1200°C for deterioration to become evident.[45])

Completely frozen microstructure was found in this work and in the literature for $Al_2O_3$[15], $BaTiO_3$[17], $Y_2O_3$[53], and $Ni_{0.2}Cu_{0.2}Zn_{0.6}Fe_2O_4$[17] in $T_2$-sintering for up to 40 h, which is enough to reach full density and is a comparable schedule used in industry, e.g., for MLCC.

In summary, new theoretical and statistical analysis of microstructure evolution during sintering has revealed a sweet spot at an intermediate density that enables highly uniform bulk, dense nanomaterials to obtain for ceramics and powder-metallurgy metals alike, and it also explains why materials of superior reliability often arise from two-step sintering. This was expressly demonstrated for a dense bulk nanocrystalline $Al_2O_3$, a ceramic known to be difficult to sinter without sintering additives, yet the achieved grain-size uniformity exceeded the theoretical "limit" predicted by Hillert 60 years ago. The success will encourage the development of other highly uniform nanocrystalline materials for structural and functional applications.

## Methods

$\alpha$-$Al_2O_3$ powders were prepared by high-energy ball milling followed by corrosion and separation.[61] $\alpha$-$Al_2O_3$ powders with varying sizes and purities (see **Table S2** for details) were obtained by adjusting the pH during the separation process. Pressed pellets of $\alpha$-$Al_2O_3$ powders were heated in air at 10°C/min to various temperatures, then either held there for a certain holding time or immediately cooled as specified in **Supplementary Information**. Two-step sintering was conducted by firstly heating the pellet at 10°C/min to 1150°C, then, without holding, immediately cooling it at 5°C/min to 1025°C and held there for 40 h. The theoretical density of $Al_2O_3$ was chosen as 3.96 g/cm$^3$. Fractured fragments of sintered pellets which underwent intergranular fracture were examined under a transmission electron microscope (TEM; FEI Tecnai G2 F30). Relatively flat fractured and thermally etched (at 50°C lower than sintering temperature for 0.5 h) surfaces of the pellets were examined under scanning electron microscope (SEM; Tescan LYRA3 XMU).

Pressed pellets of 8YSZ powders (TZ-8Y, Tosoh Co., Tokyo, Japan) were heated in air to 1280°C without holding to obtain porous sample, or to 1300°C and held there for 12 h to obtain dense sample, both with 5°C/min heating/cooling rate. Pressed pellets of 3YSZ powders (TZ-3Y-E, Tosoh Co., Tokyo, Japan) were heated in air to 1240°C without holding to obtain porous sample, or to 1300°C and held there for 12 h to obtain dense sample, both with 5°C/min heating/cooling rate. The theoretical density of 8YSZ/3YSZ was chosen as 6.08 g/cm$^3$. Polished and thermally etched (1220°C for 0.2 h for porous/dense 8YSZ and dense 3YSZ, 1150°C for 0.2 h for porous 3YSZ) surfaces were examined under SEM (Quanta 600, FEI Co.).

Mo powders were prepared via a solution combustion method[62] followed by hydrogen reduction, using ammonium molybdate hydrate $(NH_4)_6Mo_7O_{24}\cdot4H_2O$ as a precursor. Pressed pellets of Mo powders were sintered in a hydrogen atmosphere with 10°C/min heating rate and 5°C/min cooling rate. Normal sintering was conducted at 1400°C for 2 h. Two-step sintering was conducted by firstly heating the pellet to 1180°C and held for 1 h, then cooling it to 1100°C and held there for 10 h. The theoretical density of Mo was chosen as 10.2 g/cm$^3$. Relatively flat fractured surfaces of the pellets, which underwent intergranular fracture, were examined under SEM (Hitachi Uhr SU8100).

90W-10Re powders were prepared via a similar solution combustion method followed by hydrogen reduction, using ammonium metatungstate hydrate $(NH_4)_6H_2W_{12}O_{40}\cdot xH_2O$ and ammonium perrhenate $NH_4ReO_4$ as precursors with a weight ratio of W:Re=9:1. Pressed pellets of 90W-10Re powders were sintered in a hydrogen atmosphere with 10°C/min heating rate and 5°C/min cooling rate. Normal sintering was conducted at 1500°C for 2 h. Two-step sintering was conducted by firstly heating the pellet to 1200°C and held for 1 h, then cooling it to 1100°C and held there for 20 h. The theoretical density of 90W-10Re was chosen as 19.4 g/cm$^3$. Relatively flat fractured surfaces of the pellets, which underwent intergranular fracture, were examined under SEM (FEI Quanta FEG 450).

The BT samples analyzed were from a previous study.[8] As described therein, powders were prepared by a precipitation/coating method on commercial $BaTiO_3$ powders (KCM Co. Ltd., Nagoya, Japan). Pressed pellets were sintered in a reducing atmosphere ($H_2$-$N_2$-$H_2O$ gas mixture, oxygen partial pressure $10^{-13}$-$10^{-15}$ Pa) with heating schedules described in **Fig. 1** of Ref. [8]. The theoretical density of BT was chosen as 6.02 g/cm$^3$. Surfaces of the pellets were examined under SEM (MERLIN VP Compact, Carl Zeiss Corp.).

The BT-BZNT samples analyzed were from another previous study.[57] As described therein, powders were prepared by a solid-state synthesis method, and their green body tapes for multilayer capacitors were prepared by a tape casting method.

Normal sintering was conducted at 1150°C for 3 h. Two-step sintering was conducted by firstly heating the pellet to 1190°C and held for 1 min, then cooling it to 1040°C and held there for 3 h. Surfaces of the sintered pieces were examined under SEM (Supra 40/40vp, Carl Zeiss Corp.).

Average grain sizes and grain size distributions were calculated by measuring the maximum diameter (i.e., along the direction that gives the largest diameter) for $N$ grains ($N$ typically in the range of 300-600; see **Table S1-S3** for more details).

## Acknowledgements

I.W.C. and Y.D. acknowledge the support by the Department of Energy (BES grant no. DEFG02-11ER46814) and the LRSM facilities funded by the U.S. National Science Foundation (grant no. DMR-1120901) during Y.D.'s PhD research at the University of Pennsylvania that initiated the present work. Ju L., Y.D. and D.D. acknowledge the support by the U.S. Department of Energy (USDOE), Office of Energy Efficiency and Renewable Energy (EERE), Advanced Manufacturing Office (AMO) R&D Projects Emerging Research Exploration, under DOE Idaho Operations Office with contract no. DE-AC07-05ID14517. Jiangong L. and H.Y. acknowledge the support by the National Natural Science Foundation of China (51772137) and the Fundamental Research Funds for the Central Universities (lzujbky-2019-sp03). X.W. acknowledges the support by the National Key Research and Development Program of China (grant no. 2017YFB0406302) and the Key Area Research Plan of Guangdong (grant no. 2019B010937001).

**Conflict of interest**

The authors declare no conflict of interest.

# Supplementary Information

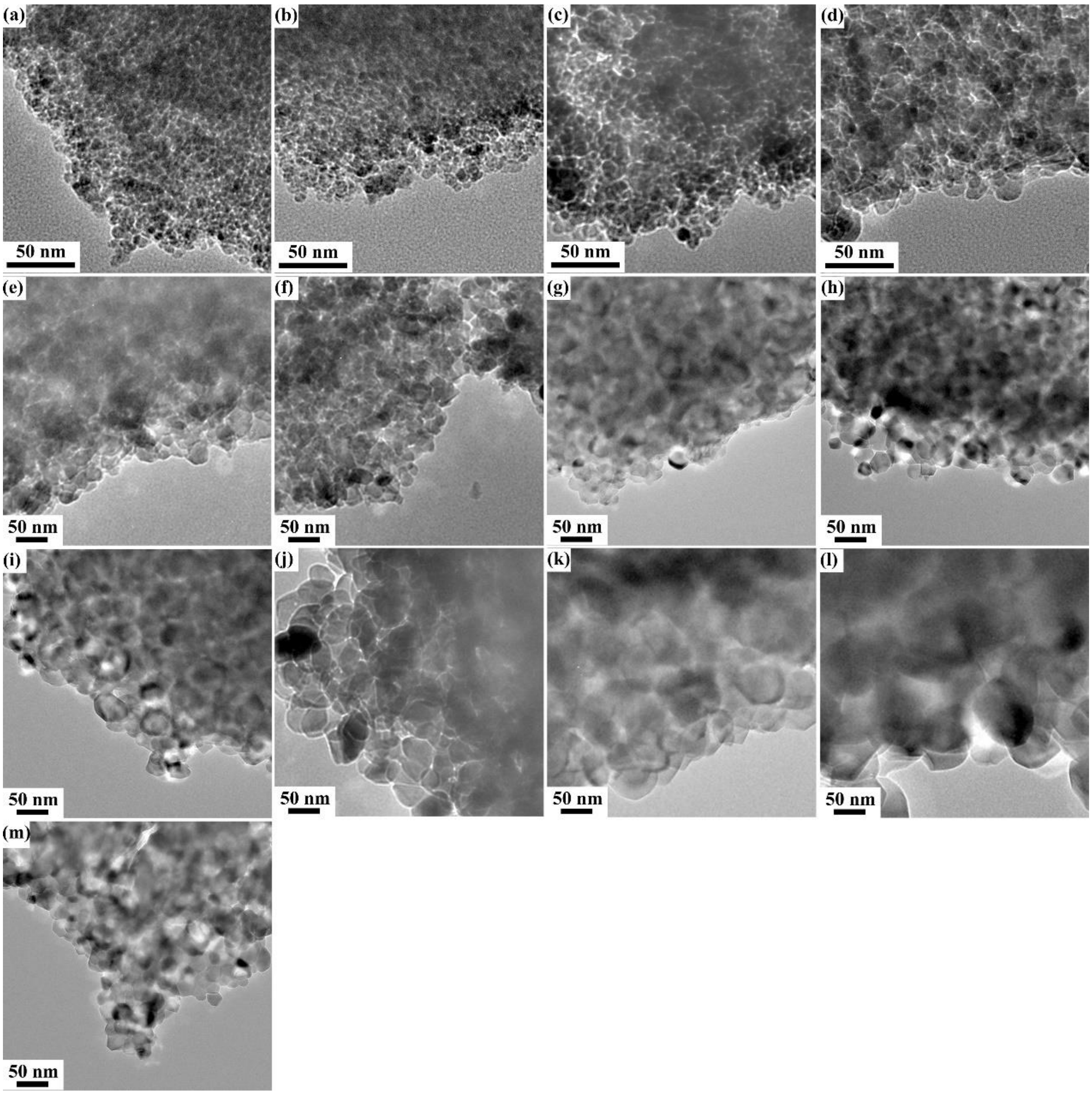

**Figure S1** TEM images of $Al_2O_3$ heated to (a) 700°C, (b) 800°C, (c) 900°C, (d) 1000°C, (e) 1050°C, (f) 1100°C, (g) 1125°C, (h) 1150°C, (i) 1175°C, (j) 1200°C, (k) 1250°C, and (l) 1300°C, all without holding. (m) Two-step sintered $Al_2O_3$ at 1150°C ($T_1$) without holding and then at 1025°C ($T_2$) for 40 h.

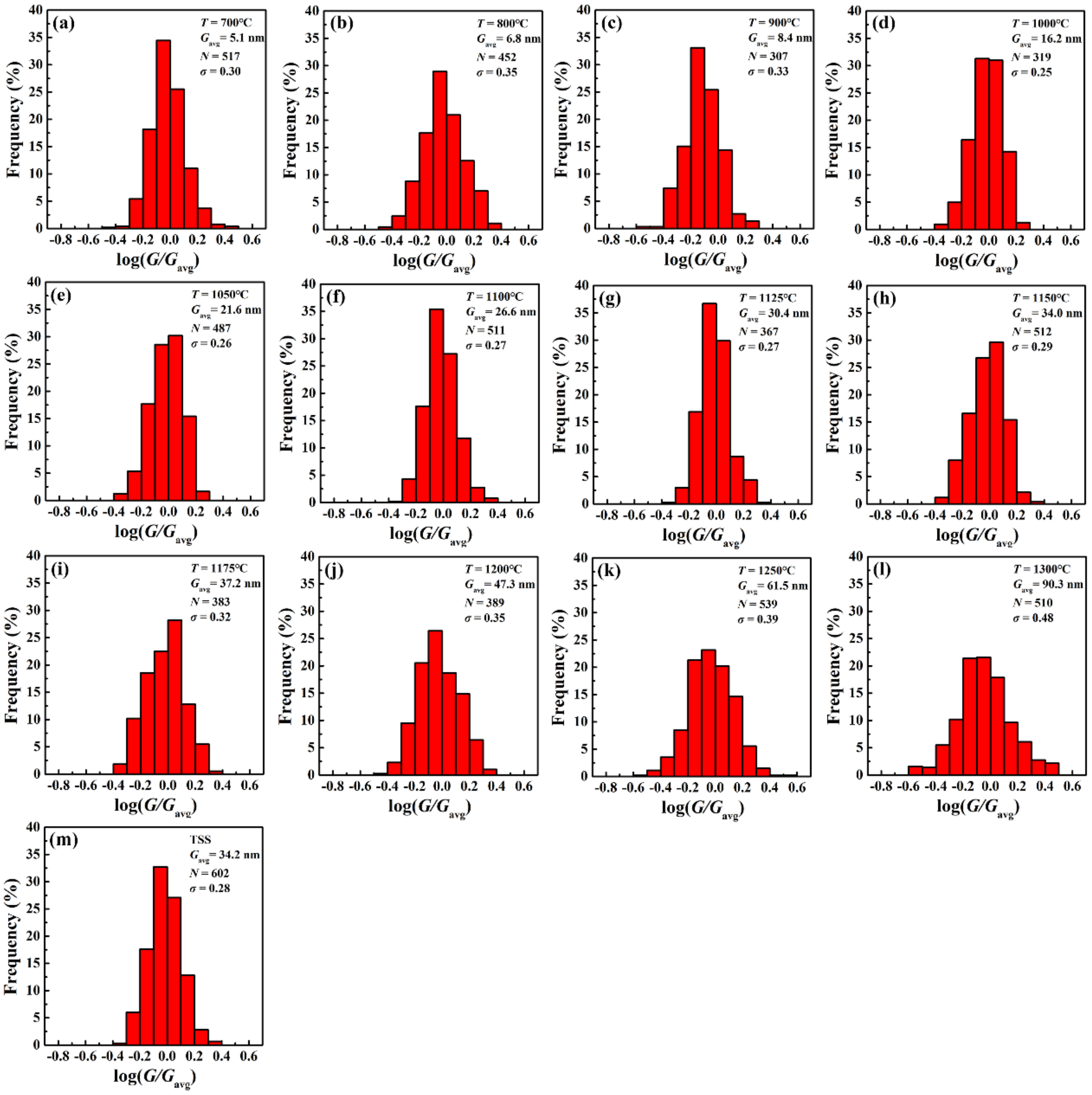

**Figure S2** Normalized grain size distribution measured from TEM images in **Fig. S1**. Also listed are $G_{avg}$, number $N$ of measured grains and $\sigma$. $Al_2O_3$ samples were sintered at (a) 700°C, (b) 800°C, (c) 900°C, (d) 1000°C, (e) 1050°C, (f) 1100°C, (g) 1125°C, (h) 1150°C, (i) 1175°C, (j) 1200°C, (k) 1250°C, and (l) 1300°C, all without holding. (m) Two-step sintered $Al_2O_3$ at 1150°C ($T_1$) without holding and then at 1025°C ($T_2$) for 40 h.

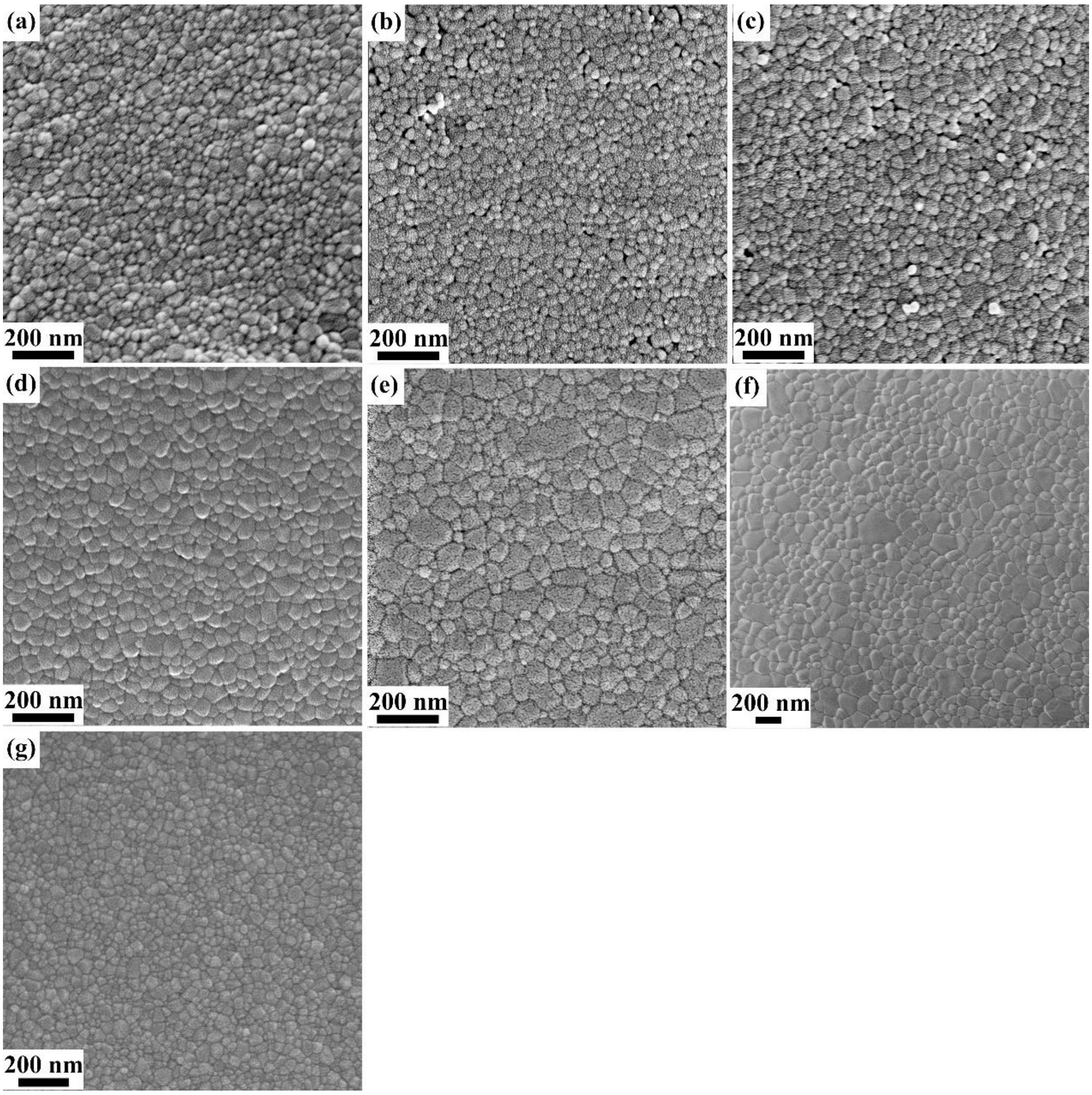

**Figure S3** SEM images of $Al_2O_3$ sintered at (a) 1125$^o$C, (b) 1150$^o$C, (c) 1175$^o$C, (d) 1200$^o$C, (e) 1250$^o$C, and (f) 1300$^o$C, all without holding. (g) Two-step sintered $Al_2O_3$ at 1150$^o$C ($T_1$) without holding and then at 1025$^o$C ($T_2$) for 40 h.

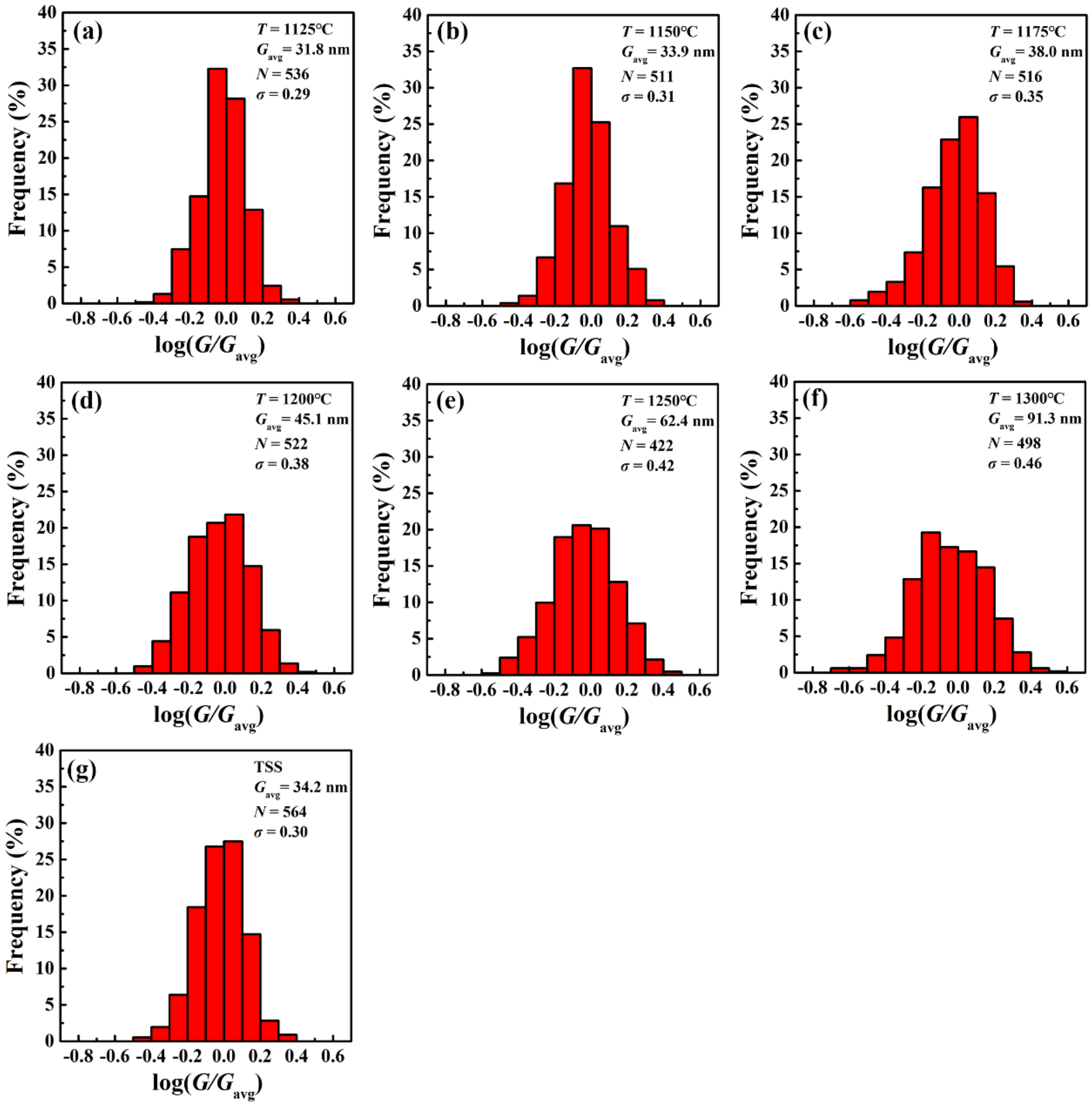

**Figure S4** Normalized grain size distribution measured from SEM images in **Fig. S3**. Also listed are $G_{avg}$, number $N$ of measured grains and $\sigma$. $Al_2O_3$ samples were sintered at (a) 1125°C, (b) 1150°C, (c) 1175°C, (d) 1200°C, (e) 1250°C, and (f) 1300°C, all without holding. (g) Two-step sintered $Al_2O_3$ at 1150°C ($T_1$) without holding and then at 1025°C ($T_2$) for 40 h.

**Table S1** Measured $G_{avg}$, $N$ and $\sigma$ for dense polycrystalline materials. Materials processed by two-step sintering in red, and by other techniques in blue.

| Materials | Symbol | $G_{avg}$ | $N$ | $\sigma$ | Reference |
|---|---|---|---|---|---|
| $Al_2O_3$ | Filled star | 385nm | 431 | 0.57 | This work |
| $Al_2O_3$ | Filled square | 0.33μm | 209 | 0.50 | Fig. 2a, Ref. 10 |
| $Al_2O_3$ | Filled square | 4.0μm | 261 | 0.69 | Fig. 3a, Ref. 11 |
| $Al_2O_3$ | Filled square | 4.4μm | 421 | 0.39 | Fig.1a, Ref. 12 |
| $Al_2O_3$ | Filled square | 31.4μm | 510 | 0.65 | Fig.1d, Ref. 12 |
| $Al_2O_3$ | Filled square | 0.24μm | 148 | 0.46 | Fig.3, Ref. 13 |
| $Al_2O_3$ | Filled square | 1.1μm | 216 | 0.48 | Fig.8b, Ref. 14 |
| $Al_2O_3$ | Filled star | 34nm | 564 | 0.30 | This work |
| $Al_2O_3$ | Filled square | 70nm | 343 | 0.36 | This work |
| $Al_2O_3$ | Filled square | 164nm | 468 | 0.39 | This work |
| $Al_2O_3$ | Filled square | 55nm | 525 | 0.34 | Fig. 6, Ref. 15 |
| $Al_2O_3$ | Filled square | 41nm | 536 | 0.36 | Fig. 7a, Ref. 16 |
| $Al_2O_3$ | Filled square | 32nm | 612 | 0.34 | Fig. 7b, Ref. 16 |
| $Al_2O_3$ | Filled square | 54nm | 530 | 0.34 | Fig. 7c, Ref. 16 |
| $BaTiO_3$ | Open square | 0.19μm | 459 | 0.47 | This work |
| $BaTiO_3$ | Open square | 0.16μm | 129 | 0.48 | Fig. 2b, Ref. 8 |
| $BaTiO_3$ | Open square | 1.2μm | 223 | 0.37 | Fig. 4f, Ref. 17 |
| $BaTiO_3$ | Open square | 0.18μm | 514 | 0.36 | This work |
| $BaTiO_3$ | Open square | 0.16μm | 127 | 0.44 | Fig. 2d, Ref. 8 |
| $BaTiO_3$ | Open square | 68nm | 532 | 0.37 | Fig. 4b, Ref. 17 |
| $BaTiO_3$ | Open square | 98nm | 357 | 0.42 | Fig. 4c, Ref. 17 |
| $BaTiO_3$ | Open square | 0.20μm | 521 | 0.38 | Fig. 4d, Ref. 17 |
| BT-BZNT | Open square | 0.66μm | 401 | 0.60 | This work |
| BT-BZNT | Open square | 0.52μm | 535 | 0.48 | This work |
| $BaZr_{0.1}Ce_{0.7}Y_{0.1}Yb_{0.1}O_{3-x}$ | Half-filled square | 0.49μm | 98 | 0.40 | Fig.3a, Ref. 18 |
| $BaZr_{0.1}Ce_{0.7}Y_{0.1}Yb_{0.1}O_{3-x}$ | Half-filled square | 0.57μm | 80 | 0.40 | Fig.3b, Ref. 18 |
| $BaZr_{0.1}Ce_{0.7}Y_{0.1}Yb_{0.1}O_{3-x}$ | Half-filled square | 0.30μm | 229 | 0.36 | Fig.3c, Ref. 18 |
| $BaZr_{0.1}Ce_{0.7}Y_{0.1}Yb_{0.1}O_{3-x}$ | Half-filled square | 1.3μm | 486 | 0.36 | Fig.5a, Ref. 19 |
| $BaZr_{0.1}Ce_{0.7}Y_{0.1}Yb_{0.1}O_{3-x}$ | Half-filled square | 2.3μm | 176 | 0.39 | Fig.5b, Ref. 19 |
| $CeO_2$ | Filled cycle | 0.67μm | 115 | 0.49 | Fig. 2a, Ref. 20 |
| $CeO_2$ | Filled cycle | 0.39μm | 119 | 0.44 | Fig. 2b, Ref. 20 |
| $CeO_2$ | Filled cycle | 16.6μm | 93 | 0.52 | Fig. 2c, Ref. 20 |
| $CeO_2$ | Filled cycle | 0.54μm | 76 | 0.42 | Fig. 2d, Ref. 20 |
| $CeO_2$ | Filled cycle | 20.9μm | 74 | 0.43 | Fig. 2e, Ref. 20 |
| $Gd_{0.1}Ce_{0.9}O_{1.95}$ | Open cycle | 0.51μm | 237 | 0.43 | Fig.1c, Ref. 21 |
| $Gd_{0.1}Ce_{0.9}O_{1.95}$ | Open cycle | 0.77μm | 231 | 0.38 | Fig.5b, Ref. 22 |
| $(K,Na,Li)(Nb,Ta,Sb)O_3$ | Half-filled cycle | 1.6μm | 190 | 0.60 | Fig.5d, Ref. 23 |
| $Li_7La_3Zr_2O_{12}$ | Filled triangle | 5.4μm | 426 | 0.49 | Fig.1b, Ref. 24 |
| $Li_7La_3Zr_2O_{12}$ | Filled triangle | 2.2μm | 332 | 0.41 | Fig.3a, Ref. 25 |
| $Li_7La_3Zr_2O_{12}$ | Filled triangle | 29.1μm | 233 | 0.47 | Fig.3b, Ref. 25 |
| $Li_7La_3Zr_2O_{12}$ | Filled triangle | 39.2μm | 144 | 0.46 | Fig.3c, Ref. 25 |
| $Li_7La_3Zr_2O_{12}$ | Filled triangle | 37.6μm | 150 | 0.47 | Fig.3d, Ref. 25 |
| $(La,Sr)(Ga,Mg)O_3$ | Open triangle | 1.8μm | 121 | 0.46 | Fig.1c, Ref. 26 |
| $(La,Sr)(Ga,Mg)O_3$ | Open triangle | 2.2μm | 131 | 0.42 | Fig.1d, Ref. 26 |
| $(La,Sr)(Ga,Mg)O_3$ | Open triangle | 2.0μm | 161 | 0.36 | Fig.2a, Ref. 27 |
| $(La,Sr)(Ga,Mg)O_3$ | Open triangle | 3.4μm | 260 | 0.39 | Fig.6c, Ref. 28 |
| $(La,Sr)(Ga,Mg)O_3$ | Open triangle | 5.0μm | 139 | 0.37 | Fig.6d, Ref. 28 |
| $Lu_3Al_5O_{12}$ | Half-filled triangle | 0.39μm | 218 | 0.38 | Fig.3a, Ref. 29 |
| MgO | Filled diamond | 72nm | 212 | 0.43 | Fig.1a, Ref. 30 |
| $MgAl_2O_4$ | Open diamond | 93nm | 145 | 0.40 | Fig.3c, Ref. 31 |
| $MgAl_2O_4$ | Open diamond | 85nm | 171 | 0.45 | Fig.3d, Ref. 31 |
| Mo | Filled pentagon | 0.89μm | 322 | 0.49 | This work |
| Mo | Filled pentagon | 0.37μm | 259 | 0.38 | This work |
| $NaNbO_3$ | Half-filled diamond | 4.8μm | 156 | 0.36 | Fig.4a, Ref. 32 |
| $(Pb,La)(Zr,Ti)O_3$ | Filled hexagon | 3.2μm | 532 | 0.48 | Fig. 16, Ref. 33 |
| (Pb,La)(Zr,Ti)$O_3$ | Filled hexagon | 1.2μm | 111 | 0.44 | Fig.18, Ref. 34 |
| $SrTiO_3$ | Open hexagon | 5.8μm | 421 | 0.68 | Fig. 3a, Ref. 35 |
| $SrTiO_3$ | Open hexagon | 4.3μm | 457 | 0.96 | Fig. 3b, Ref. 35 |
| $SrTiO_3$ | Open hexagon | 3.3μm | 445 | 0.81 | Fig. 3c, Ref. 35 |
| $SrTiO_3$ | Open hexagon | 14.0μm | 506 | 0.60 | Fig. 3d, Ref. 35 |
| $SrTiO_3$ | Open hexagon | 2.6μm | 462 | 0.78 | Fig. 5a, Ref. 35 |
| $SrTiO_3$ | Open hexagon | 3.6μm | 422 | 0.84 | Fig. 5b, Ref. 35 |
| $SrTiO_3$ | Open hexagon | 5.1μm | 470 | 0.98 | Fig. 5c, Ref. 35 |
| $SrTiO_3$ | Open hexagon | 2.2μm | 519 | 0.69 | Fig. 7a, Ref. 35 |
| $SrTiO_3$ | Open hexagon | 3.1μm | 545 | 0.75 | Fig. 7b, Ref. 35 |
| $SrTiO_3$ | Open hexagon | 3.8μm | 440 | 0.86 | Fig. 7c, Ref. 35 |
| $TiO_2$ | Half-filled hexagon | 0.13μm | 371 | 0.40 | Fig. 6, Ref. 36 |
| W | Filled pentagon | 1.8μm | 371 | 0.54 | Fig. 5a, Ref. 37 |
| W | Filled pentagon | 0.83μm | 549 | 0.41 | Fig. 5b, Ref. 37 |
| 90W-10Re | Filled pentagon | 2.2μm | 253 | 0.44 | This work |
| 90W-10Re | Filled pentagon | 0.33μm | 305 | 0.35 | This work |
| $Y_2O_3$ | Open pentagon | 4.1μm | 54 | 0.42 | Fig. 3b, Ref. 38 |
| $Y_2O_3$ | Open pentagon | 0.73μm | 85 | 0.40 | Fig. 3c, Ref. 38 |
| $Y_2O_3$ | Open pentagon | 0.23μm | 98 | 0.42 | Fig. 3d, Ref. 38 |
| $Y_2O_3$ | Open pentagon | 0.31μm | 70 | 0.45 | Fig. 3e, Ref. 38 |
| $Y_3Al_5O_{12}$ | Half-filled pentagon | 0.42μm | 71 | 0.46 | Fig.8a, Ref. 39 |
| $Y_3Al_5O_{12}$ | Half-filled pentagon | 24.5μm | 431 | 0.44 | Fig.1, Ref. 40 |
| $Y_3Fe_5O_{12}$ | Open star | 5.1μm | 235 | 0.38 | Fig.5a, Ref. 41 |
| ZnO | Half-filled star | 0.34μm | 469 | 0.37 | Fig.7b, Ref. 42 |
| 3 mol% $Y_2O_3$-$ZrO_2$ | Plus | 178nm | 529 | 0.39 | This work |
| 3 mol% $Y_2O_3$-$ZrO_2$ | Plus | 78nm | 521 | 0.32 | Fig. 6a, Ref. 43 |
| 3 mol% $Y_2O_3$-$ZrO_2$ | Plus | 0.21μm | 130 | 0.39 | Fig. 6b, Ref. 43 |
| 3 mol% $Y_2O_3$-$ZrO_2$ | Plus | 88nm | 469 | 0.43 | Fig. 6c, Ref. 43 |
| 8 mol% $Y_2O_3$-$ZrO_2$ | Cross | 1.19μm | 354 | 0.53 | This work |
| 8 mol% $Y_2O_3$-$ZrO_2$ | Cross | 0.27μm | 427 | 0.34 | Fig. 9b, Ref. 44 |
| 8 mol% $Y_2O_3$-$ZrO_2$ | Cross | 0.34μm | 523 | 0.41 | Fig. 2c, Ref. 45 |

**Table S2** Measured $G_{avg}$, $N$ and $\sigma$ for three different $\alpha$-$Al_2O_3$ powders and their two-step sintered ceramics of density $\rho$.

| $\alpha$-$Al_2O_3$ powders | | | | Two-step sintered dense $Al_2O_3$ | | | | |
|---|---|---|---|---|---|---|---|---|
| Purity (wt%) | $G_{avg}$ (nm) | $N$ | $\sigma$ | Sintering condition | $\rho$ (%) | $G_{avg}$ (nm) | $N$ | $\sigma$ |
| 99.958 | 4.7 | 1018 | 0.23 | $T_1$=1150°C for 0 h, $T_2$=1025°C for 40 h | 99.6 | 34 | 564 | 0.30 |
| 99.938 | 11.6 | 675 | 0.41 | $T_1$=1200°C for 0 h, $T_2$=1000°C for 20 h | 99.1 | 70 | 343 | 0.36 |
| 99.914 | 16.1 | 526 | 0.43 | $T_1$=1275°C for 0 h, $T_2$=1075°C for 20 h | 98.9 | 164 | 468 | 0.39 |

**Table S3** Measured $\rho$, $G_{avg}$, $N$ and $\sigma$ for Mo, 90W-10Re, BT and BT-BZNT via two-step sintering and one-step sintering.

| Materials | Sintering condition | $\rho$ (%) | $G_{avg}$ (μm) | $N$ | $\sigma$ |
|---|---|---|---|---|---|
| Mo | $T_1$=1180°C for 1 h, $T_2$=1110°C for 10 h | 98.8 | 0.37 | 259 | 0.38 |
| | 1400°C for 3 h | 97.8 | 0.89 | 322 | 0.49 |
| 90W-10Re | $T_1$=1200°C for 1 h, $T_2$=1100°C for 20 h | 98.5 | 0.33 | 305 | 0.35 |
| | 1500°C for 2 h | 97.3 | 2.2 | 253 | 0.44 |
| BT | $T_1$=1250°C for 1min, $T_2$=1050 °C for 5 h | 97.8 | 0.18 | 514 | 0.36 |
| | 1210°C for 2 h | 98.0 | 0.19 | 459 | 0.47 |
| BT-BZNT | $T_1$=1190°C for 1 min, $T_2$=1040 °C for 3 h | / | 0.52 | 535 | 0.48 |
| | 1150°C for 3 h | / | 0.66 | 401 | 0.60 |